\documentclass[showkeys, showpacs, aps,prl, preprint, superscriptaddress]{revtex4}
\usepackage{color,graphicx}% Include figure files
\usepackage{dcolumn}% Align table columns on decimal point
\usepackage{bm}% bold math
\usepackage{amsmath}
\usepackage{amssymb}
\usepackage{latexsym}
\usepackage{epsfig}
\usepackage{amsbsy}
\usepackage{array}
\usepackage{amssymb}
\usepackage{setspace}
\usepackage{bm}

\def\sint{\ifmmode{- \!\!\!\!\!\! \int}
    \else{\hbox{$- \!\!\!\! \int \ $}}\fi}

\begin{document}

%\preprint{Advanced Materials}

\title{Spin mixing conductance at a well-controlled platinum/yttrium iron garnet interface}% Force line breaks with \\

\author{Z. Qiu\footnote{Author to whom correspondence should be
addressed; electronic mail: qiuzy@imr.tohoku.ac.jp}}
\affiliation{WPI Advanced Institute for Materials Research,
Tohoku University, Sendai 980-8577, Japan}
\author{K. Ando}
\affiliation{Institute for Materials Research, Tohoku University, Sendai 980-8577, Japan}
\author{K. Uchida}
\affiliation{Institute for Materials Research, Tohoku University, Sendai 980-8577, Japan}
\affiliation{PRESTO, Japan Science and Technology Agency, Saitama 332-0012, Japan}
\author{Y. Kajiwara}
\affiliation{Institute for Materials Research, Tohoku University, Sendai 980-8577, Japan}
\author{R. Takahashi}
\affiliation{Institute for Materials Research, Tohoku University, Sendai 980-8577, Japan}
\author{H. Nakayama}
\affiliation{Institute for Materials Research, Tohoku University, Sendai 980-8577, Japan}
\author{T. An}
\affiliation{Institute for Materials Research, Tohoku University, Sendai 980-8577, Japan}
\author{Y. Fujikawa}
\affiliation{Institute for Materials Research, Tohoku University, Sendai 980-8577, Japan}
\author{E. Saitoh}
\affiliation{WPI Advanced Institute for Materials Research, Tohoku University, Sendai 980-8577, Japan}
\affiliation{Institute for Materials Research, Tohoku University, Sendai 980-8577, Japan}
\affiliation{CREST, Japan Science and Technology Agency, Tokyo 102-0076, Japan}
\affiliation{Advanced Science Research Center, Japan Atomic Energy Agency, Tokai 319-1195, Japan}

\date{\today}% It is always \today, today,

\begin{abstract}
A platinum (Pt)/yttrium iron garnet (YIG) bilayer system with a well-controlled interface has been developed; spin mixing conductance at the Pt/YIG interface has been studied. Crystal perfection at the interface is experimentally demonstrated to contribute to large spin mixing conductance. The spin mixing conductance is obtained to be $1.3\times10^{18}~\rm{m^{-2}}$ at the well-controlled Pt/YIG interface, which is close to a theoretical prediction. 
\end{abstract}

\pacs{75.40.Gb, 72.25.-b, 75.30.Ds, 75.76.+j}

%\pacs{72.25.Pn}% PACS, the Physics and Astronomy
\keywords{spin pumping, spin mixing conductance, spintronics, yttrium iron garnet}%Use showkeys class option if keyword

\maketitle

%kajiwara实验证实电讯号可以转换为自旋流,提示了一中新的信息传递形式,但是自旋流的传输效率仍然是个问题 
Spin pumping enables a generation of an electric signal in the form of a spin current even in an insulator at room temperature \cite{kajiwara nature}, which encourages ef\mbox{}forts to develop a new class of insulator spintronics \cite{kajiwara nature, Uchida Spinseebeck2, Qiu ITO, Wu mixing, Magnetron detection,Wu mixingAuYIG2, Azevedo YIGPt mixing conductance}. %Since a spin current in an insulator is free from an accompanying charge current, it can be transmitted with a minimal dissipation \cite{kajiwara nature, Uchida Spinseebeck2}. 
The efficiency of the spin pumping is described in terms of spin mixing conductance at an interface. Improvement of spin mixing conductance is important for the development of spintronics devices \cite{kajiwara nature, Uchida Spinseebeck2, Qiu ITO, Wu mixing, Magnetron detection,Wu mixingAuYIG2, Azevedo YIGPt mixing conductance, Dazhi Bi, Ando nature semiconductor, Ando semiconductor, Uchida Spinseebeck1, others Spinseebeck, Uchida Spinseebeck3}. On the other hand, atomic structures at a metal/insulator interface have not been successfully observed or controlled thus far, and such situations can be the reason of scatter in the experimentally obtained values of spin mixing conductance. 

To overcome this problem, we prepared a bilayer spin pumping system composed of a well-controlled single-crystalline yttrium iron garnet (YIG) surface covered with a platinum (Pt) film, and we studied the spin mixing conductance at the Pt/YIG interface in this study. This interface has been studied extensively because of the extremely small magnetic damping of YIG \cite{YIGdamping1, YIGdamping2, YIGdamping3} and an effective spin current detector of Pt in the sense of the inverse spin Hall ef\mbox{}fect (ISHE) \cite{ISHE Saitoh, ISHE Azevedo}. %We demonstrate experimentally that crystal perfection at the interface contributes to a large spin mixing conductance. At a well-controlled Pt/YIG interface, we obtained the value of spin mixing conductance as $1.3\times10^{18}~\rm{m^{-2}}$, which is close to a theoretical prediction \cite{AgYIG theory}.

%spin pumping的效率由mixing支配
%mixing的机理仍然不为所知. 
Spin mixing conductance refers to the ef\mbox{}ficiency with which the spin currents across at the Pt/YIG interface are generated. In a model of spin pumping \cite{bauer SP model,bauer SP model2}, the DC component of the generated spin current density, $j_s^0$, in the Pt layer at the interface can be expressed as
\begin{eqnarray}
{ j }_{ s }^{0}=\frac { \omega  }{ 2\pi  } \int _{ 0 }^{ { 2\pi  }/{ \omega  } }{ \frac { \hslash  }{ 4\pi  } { g }_{ r }^{ \uparrow \downarrow  }\frac { 1 }{ { M }_{ s }^{ 2 } } { \left[ \boldsymbol{\rm{M}}(t)\times \frac { d\boldsymbol{\rm{M}}(t) }{ dt }  \right]  }_{ z }dt }  ,
\label{eq:mixing conductance}
\end{eqnarray}
where $\omega $, $\hslash$, $g_r^{\uparrow\downarrow}$, $M_s$, and $\boldsymbol{\rm{M}}(t)$ are the angular frequency of magnetization precession, the Dirac constant, the real part of spin mixing conductance at the interface, the saturation magnetization of YIG, and the precessing magnetization in YIG, respectively. The expression $ { \left[ \boldsymbol{\rm{M}}(t)\times { d\boldsymbol{\rm{M}}(t) }/{ dt }  \right]  }_{ z }$ is the $z$ component of $\boldsymbol{\rm{M}}(t)\times { d\boldsymbol{\rm{M}}(t) }/{ dt }$. The $z$ axis is directed along the magnetization-precession axis. The spin current density $j_s^0$ can be detected as a voltage signal $V_{\rm{ISHE}}$ via ISHE in the Pt layer, which enables the experimental estimation of the spin mixing conductance $g_r^{\uparrow\downarrow}$ \cite{Ando metal,Ando metal2,Nakayama Pt}. %In this work, a well-controlled Pt/single-crystalline YIG interface was developed to study the mixing conductance at the Pt/YIG interface.

YIG films used in this work were grown on (111) gadolinium gallium garnet substrates by using a liquid phase epitaxy method in $\rm{PbO-B_2O_3}$ flux at the temperature of 1,210 K. The thickness of the YIG films was about 4.5 $\rm{\mu m}$. We preformed annealing with an oxygen pressure of ~ $5\times10^{-5}$ Torr (Fig.~\ref{fig:Graph1} (a)) to improve the crystal perfection of the YIG surface, or bombarded the YIG surface with accelerated ion beams to create an obvious amorphous layer (Insets to Fig.~\ref{fig:Graph3}). 10 nm-thick Pt films were deposited on those YIG films using a pulse laser deposition system at room temperature in the same vacuum chamber. 

ref\mbox{}lection high energy electron dif\mbox{}fraction (RHEED) was used to observe the surface structure of the YIG film during the annealing.~The Pt/YIG interface structure was characterized by a high-resolution transmission electron microscopy (TEM). The crystalline characterization was carried out by using an X-ray dif\mbox{}fraction (XRD) system and TEM. The lattice constant was calculated from the XRD and electron dif\mbox{}fraction patterns. The saturation magnetization $4\pi M_s$ and the gyromagnetic ratio, $\gamma$ were estimated by using the method described in \cite{Ando metal, Ando metal2}. Electrical conductivity $\sigma_{N}$ of the Pt film was measured by using a four-probe method. 

%通过一个ESR系统,上述试样的Spin pumping效果被研究
Spin pumping was measured by using an electron spin resonance system. The sample was placed near the center of a $\rm{TE_{011}}$ cavity, where the magnetic field component of the microwave mode is maximized and the electric-field component is minimized. A microwave with the frequency of $f$=9.44 GHz was excited in the cavity. A static magnetic field $H$ was applied to the sample. The microwave absorption intensity $I$ and the ISHE voltage signal $V_{\rm{ISHE}}$ between the two electrodes attached to the Pt layer were measured at room temperature. The sign of $V_{\rm{ISHE}}$ reverses when the direction of the external magnetic field is reversed (Fig.~\ref{fig:Graph2} (b)), which is consistent with the spin pumping \cite{bauer SP model,bauer SP model2}. Both $I$ and $V_{\rm{ISHE}}$ spectra are fitted by using multi-Lorentz functions (Fig.~\ref{fig:Graph2}(b), (c)). The fitted peak positions and the full width at half maxima (FWHMs) of the $V_{\rm{ISHE}}$ spectrum are in a good agreement with the values obtained for the $I$ spectrum.

 %完全控制的PM/FM界面被制备
A change in the surface structure was observed in situ by RHEED using an electron beam along the [110] direction of the YIG surface.  Dif\mbox{}fraction patterns were recorded with a digital camera during the annealing (Figs.~\ref{fig:Graph1}(b), (c), and (d)). Several halo rings along with the weak streak patterns were observed in the RHEED patterns at the beginning of annealing (Fig.~\ref{fig:Graph1}(b)), which suggests that a thin polycrystalline layer exists on the YIG surface. For a higher annealing temperature, the halo rings disappeared as a result of the surface reintegration (Fig.~\ref{fig:Graph1}(c)). After being annealed at 1073 K for 2 hours, sharp streak patterns appeared, and were prominent even in the first Laue zone (Fig.~\ref{fig:Graph1}(d)). This implies that the annealed YIG has an atomically smooth surface with a single crystalline structure. 

%由TEM观察知道,YIG/Fe界面的结晶性良好
The cross-sectional structure of the prepared Pt/YIG interface was observed by a high-resolution TEM (Fig.~\ref{fig:Graph1}(e)). the crystal perfection of the YIG surface was found to be well kept; the Pt/YIG interface was clear and a lattice-like contrast was observed even at the first several atomic layers at the YIG side in (Fig.~\ref{fig:Graph1}(e)). The YIG layer has a cubic garnet structure, whose unit cell is shown in Fig.~\ref{fig:Graph1}(f). A projection of the unit cell along the [112] direction is shown in Fig.~\ref{fig:Graph1}(g). The lattice constant of the YIG film, determined by both ED and XRD, was 12.376 \AA, which is in good agreement with the value found in literature \cite{YIGdamping1}. On the other hand, the Pt layer showed a typical multi-crystalline structure without a preferred orientation. The lattice constant of the Pt layer was 3.92 \AA, which is also in good agreement with the value reported in literature value \cite{Pt lattice constant}. Such a well-controlled and well-defined Pt/YIG interface is a good reference for comparison with interfaces discussed in theoretical works. 

%本工作首次,显示了界面的结晶状态对mixing支配性的影响
Figure~\ref{fig:Graph3} shows ISHE voltage signal $V_{\rm{ISHE}}$ at a 20 mW microwave power for two Pt/YIG samples with dif\mbox{}ferent interface structures, which are shown in the insets to Fig.~\ref{fig:Graph3}. One sample has a well-controlled Pt/YIG interface, while a nano-scale amorphous layer was formed at the Pt/YIG interface in another sample. The ISHE voltage signal $V_{\rm{ISHE}}$ observed in the sample with a bad interface decreased markedly compared to the ISHE voltage signal of the sample with a well-controlled interface. Because the Pt layers of the two samples were prepared under the same conditions with the same thickness, the decrease in $V_{\rm{ISHE}}$ can be attributed to the Pt/YIG interface condition; the thin amorphous layer blocks the efficient spin exchange between Pt and YIG and decreases spin mixing conductance ${ g }_{ r }^{ \uparrow \downarrow  }$ (Eq.~\ref{eq:mixing conductance}). This result suggests that a better quality interface is necessary for larger spin mixing conductance. 

%为计算mixing,薄膜的体属性应该被考虑
Figure~\ref{fig:Graph2} shows the observed spectra of microwave absorption intensity $I$ and ISHE voltage signal $V_{\rm{ISHE}}$ at 10 mW microwave power of a sample with a well-controlled Pt/YIG interface.
%Both $I$ and $V_{\rm{ISHE}}$ spectra are fitted by using multi-Lorentz functions. 
One advantage of a good interface structure is that it allows, not only $I$ but also $V_{\rm{ISHE}}$ spectra, a clear separation of spectral contributions from dif\mbox{}ferent spin wave modes, e.g. magnetostatic surface spin wave, backward volume magnetostatic spin wave, and ferromagnetic resonance (FMR) modes (Figs.~\ref{fig:Graph2}(a) and (b)).
%The fitted peak positions and the full width at half maxima (FWHMs) of the $V_{\rm{ISHE}}$ spectrum are in a good agreement with the values obtained for the $I$ spectrum.
%In order to estimate the spin mixing conductance, the Gilbert damping constant $\alpha$ of the YIG film and the spin Hall angle $\theta _{\rm{SHE}}$ of the Pt film should be estimated precisely. 
The Gibert damping constant $\alpha$  can be estimated from FWHM $W$ of a FMR absorption peak as $\alpha=\gamma W/2\omega$, where $\gamma$ is the gyromagnetic ratio. The value of $\alpha$ was calculated to be $1.7\times10^{-3}$ from $W=1.1~\rm{mT}$ for the FMR mode. 

%mixing被计算为1.3*1018
The spin current density, $j_s^0$ gererated by the FMR excitation, is estimated to be $2.9\times 10^{-10}~\rm{Jm^{-2}}$ at 10 mW microwave power in the Pt/YIG system by using the following equation \cite{Ando metal2, Nakayama Pt}:
\begin{eqnarray}
V_{ \rm{ISHE} }=\frac { w{ \theta  }_{\rm{SHE} }{ \lambda  }_{ N }\tanh { \left( { { d }_{ N } }/{ 2{ \lambda  }_{ N } } \right)  }  }{ { d }_{ N }{ \sigma  }_{ N } } \left( \frac { 2e }{ \hbar  }  \right) { j }_{ s }^{ 0 } .
\label{eq:V_js}
\end{eqnarray}
Here,  $w$, $d_N$, $\sigma _N$, $\theta _{\rm{SHE}}$, and $\lambda _N$ are, respectively, the distance between the two electrodes of the sample, the thickness, the electric conductivity, the spin Hall angle, and the spin dif\mbox{}fusion length of Pt. The quantities $w$, $d_N$, and $\sigma _N$ were measured to be 1.5 mm, 10 nm, and $2.2\times 10^{6} ~\rm{\Omega^{-1} m^{-1}}$. We used a literature value to set $\lambda _N=$7.7 nm \cite{Nakayama Pt}. The spin Hall angle $\theta _{\rm{SHE}}$ of the Pt film was estimated to be 0.012 by using a Pt/permalloy bilayer spin pumping system, in which the Pt layer was prepared under the same conditions as the Pt/YIG samples and had the same thickness as them. \cite{Ando metal2, Nakayama Pt}. The $V_{\rm{ISHE}}$ induced by the FMR excitation was estimated to be $5.86~ \rm{\mu V}$ at 10 mW microwave power (Fig.~\ref{fig:Graph2}(b)). 

By using the Landau-Lifshitz-Gilbert equation, Eq.~\ref{eq:mixing conductance} yields 
\begin{eqnarray}
{ j }_{ s }^{ 0 }=\frac { { g }_{ r }^{ \uparrow \downarrow  }{ \gamma  }^{ 2 }{ h }^{ 2 }\hslash \left[ 4\pi { M }_{ s }\gamma +\sqrt { { \left( 4\pi { M }_{ s } \right)  }^{ 2 }{ \gamma  }^{ 2 }+4{ \omega  }^{ 2 } }  \right]  }{ 8\pi { \alpha  }^{ 2 }\left[ { \left( 4\pi { M }_{ s } \right)  }^{ 2 }{ \gamma  }^{ 2 }+4{ \omega  }^{ 2 } \right]  } ,
\label{eq:mixing conductance2}
\end{eqnarray}
where $h$ is the amplitude of the microwave magnetic field. In this work, $4\pi M_s=0.171$ T, $\gamma=1.76\times10^{11}~ \rm{T^{-1}s^{-1}}$, $h=3.58\times 10^{-5} ~\rm{T}$, and $\omega=5.93\times 10^{10}~\rm{s^{-1}}$. We estimated the spin mixing conductance  as ${ g }_{ r }^{ \uparrow \downarrow  }\approx1.3\times10^{18}~\rm{m^{-2}}$ at the Pt/YIG interface, which is close to a theoretical prediction by Jia \textit{et al.}\cite{AgYIG theory}.
 % Although, Jia \textit{et al.} predicted that a different crystal direction and/or a selected interface atomic layer of YIG may increase the spin mixing conductance little more, because of the magnetocrystalline anisotropy and/or magnetic moment density changing at the interface.
%The value of the spin mixing conductance obtained in this work may be close to the upper limit of spin mixing conductance at the Pt/YIG interface. 

%However, there may be not an order of magnitude difference of spin mixing conductance as the prediction by Bauer at al.\cite{AgYIG theory}.

%这个mixing在现有的Pt/YIG界面中是最大的,而且接近理论计算
%追求高的mixing良好的界面结晶性是重要的必须条件
%mixing仍然有提高的可能性,by结晶取向以及结晶面控制等

In summary, we prepared a single crystalline YIG surface covered with a Pt film and studied  its spin mixing conductance, which governs the spin pumping ef\mbox{}f\mbox{}iciency at the interface. Crystal perfection is experimentally demonstrated to contribute to a large spin mixing conductance. With a well-controlled interface, the spin mixing conductance of a Pt/YIG interface reached the value of $1.3\times10^{18}~\rm{m^{-2}}$, which is close to a theoretical prediction. This work provides an explicit guideline for pursuing large spin mixing conductance at metal/insulator interfaces.

%\section{\textsl{Experimental}}

This work was supported by Fundamental Research Grants from CREST-JST ‘‘Creation of Nanosystems with Novel Functions through Process Integration’’; NEXT from the cabinet of\mbox{}fice, Japan, a Grant-in-Aid for Scientific Research (A) (21244058); PRESTO-JST ''Phase Interfaces for Highly Ef\mbox{}ficient Energy Utilization''  all from MEXT, Japan. 

%%%%%%%%%%%%%%%%%%%%%%%%%%%%%%%%%%%%%%%%%%%
%因为当时都是从金属到金属的注入，所以把mixing搞大的兴趣也不大，因为反正能直接通电流嘛
%但是现在有个YIG了，发现YIG能传输电信号了%那么这个绝缘体电路就有戏了%那么我们怎么才能实现呢？就得有有效的注入
%这时，Bauer等人已经算了  真的可以很大很大很大啊！
%但是，实验上，却不是很大。。。怎么回事儿呢？
%我们现在，首次发现，界面好了就大了
%比随便乱做做大一个量级啊
%那么就搞明白了，要搞大mixing，就要做好界面
%以后mixing的问题就是界面的问题
%大家以后也清楚了，有木有希望搞出绝缘体电路，就指望界面了。

%%%%%%%%%%%%%%%%%%%%%%%%%%%%%%%%%%%%%%%%%%%%
%做为一种被广泛研究的自旋流传输现象，spin pumping可以向不同的材料中注入自旋流。
%自旋流的注入效率由mixing conductance支配。
%mixing condutance对界面结晶性敏感
%我们通过热处理改变，YIG界面，RHEED
%TEM像，PtYIG界面好，直接理论比较可能

%\bibliography{Multiferroic_20070522}% Produces the bibliography via BibTeX.

\newpage

A list of figures

\begin{enumerate}

\item   (a) The annealing process for YIG films. (b), (c), and (d) are the RHEED patterns of a YIG surface at dif\mbox{}ferent stages of the annealing. (e) A cross-sectional high-resolution TEM image of a Pt/YIG interface, of which YIG was annealed as the process shown in (a). (f) The unit cell of the cubic garnet  (space group Ia3d). (g) A projection of the cubic  garnet unit cell along the [210] direction.
\label{fig:Graph1}

\item The external magnetic field $H$ dependence of the inverse spin Hall voltage signal $V_{ \rm{ISHE} }$ of Pt/YIG samples at 20 mW microwave power. Cross sectional images by TEM are shown as insets. The upper sample with a well-controlled Pt/YIG interface, while a nano-scale amorphous layer was formed at the Pt/YIG interface in the lower sample.
\label{fig:Graph3}

\item (a) The external magnetic field $H$ dependence of the microwave absorption intensity $I$ and the multi-Lorentz fitting results of a Pt/YIG sample with a well-controlled Pt/YIG interface at 10 mW microwave power. (b) The external magnetic field $H$ dependence of the inverse spin Hall voltage $V_{ \rm{ISHE} }$. Insets to (b) are schematic illustrations of the spin pumping experimental setting for $+H$ and $-H$, of which the external magnetic fields are in the opposite direction. The multi-Lorentz fitting result of $V_{ \rm{ISHE} }$ spectrum for $+H$ is shown.
\label{fig:Graph2}

\end{enumerate}

\newpage
\begin{figure}
\centering
\begin{minipage}[b]{0.5\textwidth}
\centering
\includegraphics[width=3.5in]{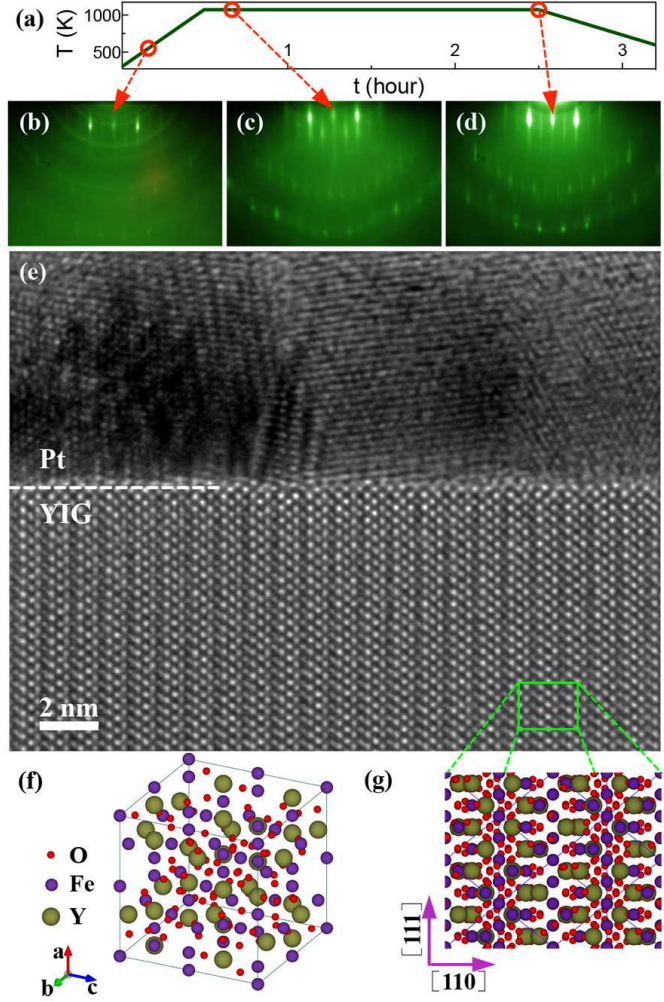}
\end{minipage}
\caption{ (a) The annealing process for YIG films. (b), (c), and (d) are the RHEED patterns of a YIG surface at dif\mbox{}ferent stages of the annealing. (e) A cross-sectional high-resolution TEM image of a Pt/YIG interface, of which YIG was annealed as the process shown in (a). (f) The unit cell of the cubic garnet  (space group Ia3d). (g) A projection of the cubic  garnet unit cell along the [210] direction.
\label{fig:Graph1}}
\end{figure}

\begin{figure}
\centering
\begin{minipage}[b]{0.5\textwidth}
\centering
\includegraphics[width=3in]{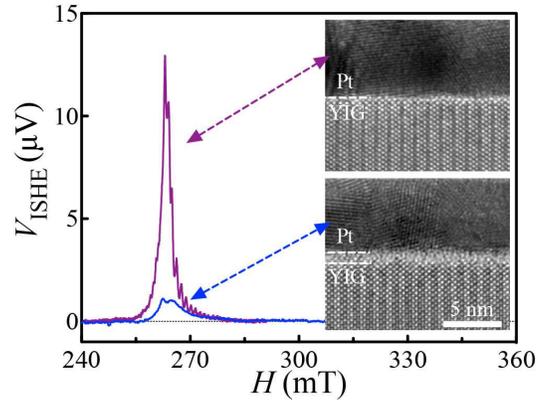}
\end{minipage}
\caption{The external magnetic field $H$ dependence of the inverse spin Hall voltage signal $V_{ \rm{ISHE} }$ of Pt/YIG samples at 20 mW microwave power. Cross sectional images by TEM are shown as insets. The upper sample with a well-controlled Pt/YIG interface, while a nano-scale amorphous layer was formed at the Pt/YIG interface in the lower sample.
\label{fig:Graph3}}
\end{figure}

\begin{figure}
\centering
\begin{minipage}[b]{0.5\textwidth}
\centering
\includegraphics[width=3in]{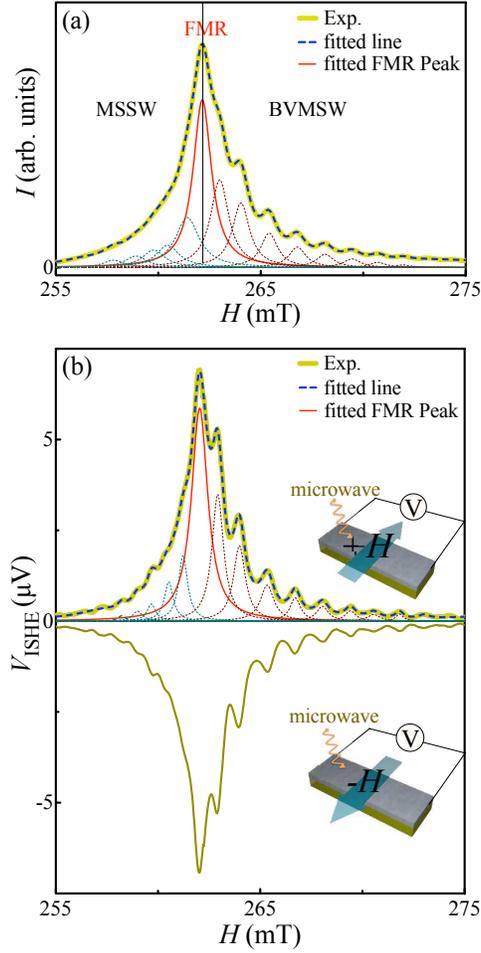}
\end{minipage}
\caption{(a) The external magnetic field $H$ dependence of the microwave absorption intensity $I$ and the multi-Lorentz fitting results of a Pt/YIG sample with a well-controlled Pt/YIG interface at 10 mW microwave power. (b) The external magnetic field $H$ dependence of the inverse spin Hall voltage $V_{ \rm{ISHE} }$. Insets to (b) are schematic illustrations of the spin pumping experimental setting for $+H$ and $-H$, of which the external magnetic fields are in the opposite direction. The multi-Lorentz fitting result of $V_{ \rm{ISHE} }$ spectrum for $+H$ is shown.
\label{fig:Graph2}}
\end{figure}

\end{document}